\documentclass[12pt]{article}
\usepackage[letterpaper,margin=1in]{geometry}
\usepackage{caption} 
\usepackage{setspace}
\usepackage{graphicx} 

\usepackage{cite}

\usepackage{textcomp}
\usepackage{amsmath}

\usepackage[dvipsnames]{xcolor}

\usepackage{xr}
\usepackage{hyperref}

\usepackage{textgreek}
\usepackage{graphicx}
\makeatletter
\newcommand*{\addFileDependency}[1]{
  \typeout{(#1)}
  \@addtofilelist{#1}
  \IfFileExists{#1}{}{\typeout{No file #1.}}
}
\makeatother

\newcommand*{\myexternaldocument}[1]{%
    \externaldocument{#1}%
    \addFileDependency{#1.tex}%
    \addFileDependency{#1.aux}%
}

\myexternaldocument{SI_final}

\usepackage{authblk}

\author[1]{Farkhad Maksudov}
\affil[1]{Department of Chemistry, The University of Texas at Austin, 105 E 24th Street, Stop A5300, Austin, TX 78712, USA.}
\author[2]{Mauro L. Mugnai}
\affil[2]{Institute of Soft Matter Synthesis and Metrology, Georgetown University, Washington DC 20057, USA.}
\author[1,3]{Laura Dominguez}
\affil[3]{Departamento de Fisicoquímica, Facultad de Química, Universidad Nacional Autónoma de México, Coyoacán, CDMX, Mexico}
\author[1,4]{Dmitrii Makarov}
\affil[4]{Oden Institute for Computational Engineering and Sciences, University of Texas at Austin, Austin, Texas 78712, United States}
\author[*,1]{D. Thirumalai}
\affil[ ]{Email: dave.thirumalai@gmail.com}

\title{Droplet growth, Ostwald's rule, and emergence of order in Fused in Sarcoma}

\date{}

\begin{document}

\maketitle

\section*{Abstract}

The low complexity domain of Fused in Sarcoma (FUS-LC consisting of 214 residues) undergoes phase separation, resulting in a dense liquid-like phase that forms early and slowly matures to reach ordered gel-like state on long time scales. Upon maturation, core-1, comprising of the 57 residues (39-95) in the N-terminus become structured, resulting in the formation of a non-polymorphic fibril. The truncated FUS-LC-C (residues 110-214) construct forms a fibril in which core-2 (residues 112-150) adopts a $\beta$-sheet structure.  Using coarse-grained monomer SOP-IDP model simulations of FUS-LC,  we predict that residues 155-190 in the C-terminal (core-3) form rapidly, followed by core-2, and finally core-1.  The time scale of formation of the cores and their stabilities are inversely correlated, as anticipated by the Ostwald's rule of stages. Unbiased multichain simulations show that the chemical potentials in the two phases are equal and the calculated densities of the dense and dilute phases are in agreement with experiments.  The dense phase, which forms by a nucleation mechanism, coarsens over time by a process that is reminiscent of Ostwald ripening. AlphaFold predictions of the core-3 structure and the simulations  show that $\beta$-strand emerges in the core-3 region early during the droplet formation, and drives the initiation of FUS-LC assembly. The techniques  introduced here are general and could be used to probe  assembly of other IDPs such as TDP-43, which shares many features with FUS-LC.

\newpage
\section*{Introduction}
Phase separation in polymers, resulting in the coexistence of dense and dilute phases, is a well known phenomenon in synthetic \cite{Flory53PolymerBook} and biological molecules \cite{Overbeek57JCellularCompPhys}.  Evidence of phase separation  in a mixture of glycoproteins and polysaccharides, and gelatin existed nearly a century ago  \cite{Jong29ProcKoninkl}, which coined the term ``complex coacervation" to describe this phenomenon). The dense phase consisted of polymers and water and the dispersed low density phase was a solution containing one or more polymers.  However, only with the demonstration that phase separation is an important mechanism of cellular organization there has been a great interest in its relevance in biology and biophysics~\cite{Brangwynne09Science,Alberti19Cell,Banani17NatRevMolCellBiol,Choi20ARB}. Several studies have shown that besides proteins, RNA in complex with proteins and chromatin also phase separate spontaneously both \textit{in vivo} and \textit{in vitro} \cite{Rippe22CSHBiol,Roden21NatRevMolCellBiol,Trcek20MolCell}. In many cases, proteins, typically intrinsically disordered proteins (IDPs) or proteins with disordered regions, could phase separate, at least in \textit{in vitro}. Although the vast number of experimental~\cite{Alberti21NatRevMolCellBiol}, theoretical \cite{Thirumalai24Biopolymers}, and computational studies ~\cite{Shea21COSB,Ranganathan23PNAS,Ranganathan22BJ,Joseph21NatComSci,Dignon20ARPC,tesei2021accurate,Wang25NatComm} have  shed light on the molecular basis of phase separation, many unanswered questions remain. Here, we investigate the mechanism of droplet growth during phase separation in Fused in Sarcoma (FUS), an RNA-binding protein, using computer simulations. Our focus is on the low-complexity (LC) N-terminal domain of FUS (referred to as FUS-LC), which is critical for its ability to undergo phase separation and to form fibril that bind to hydrogels \cite{core1Tycko,core3_McKnight}. 

The 214 residue long  FUS-LC (the see the sequence in Fig. \ref{fig:True_MFPT}A), which forms non-polymorphic fibrils, has several  unusual attributes that are also shared by other IDPs, such as TDP43 and hnRNPA1. A number of experimental studies  \cite{patel2015liquid,core3_McKnight}, summarized previously \cite{Thirumalai24Biopolymers}, have shown that FUS-LC and related constructs undergo phase separation under nominal conditions (room temperature and neutral pH).  
(1) Solid state NMR experiments showed that, on very long time scales,  FUS-LC forms a non-polymorphic fibril in which residues 39-95 (referred to as core-1) form a cross-$\beta$ structure with an S-bend topology \cite{core1Tycko}. In contrast, FUS-LC-C (residues 111-214) also forms a non-polymorphic structure but with a  protolfilament consisting of  a dimer (core-2) with each monomer resembling a U-bend in which only  residues 112-150 are structured. The two ordered cores structures do not coexist in FUS-LC even though their free energies are similar \cite{core2Tycko,kumar2021sequence}. 
(2) A more recent study demonstrated \cite{core3_McKnight} that a C-terminal region, specifically residues 155 to 190 (core-3), plays a crucial role in hydrogel binding, which means that it is structured~\cite{Kato22RNA}. Using a combination of tyrosine-to-serine mutations and systematic deletion of residues from the C-terminal ends of FUS-LC, it was found that mutations in the C-terminal region significantly reduced the ability of FUS-LC to bind to preformed hydrogels \cite{core3_McKnight}. Surprisingly, even after deleting a substantial portion of the N-terminal residues (1-61) that includes part of the core-1 residues (39-61),  hydrogel binding was not abolished. These findings indicate that FUS self-assembly must involve at least two distinct interaction regions, raising the question of how these regions interact to regulate FUS assembly. 
(3) Deletion of most of the residues in core-2 retains the ability to bind to the hydrogel. 
The results, summarized in (2) and (3), suggest that the FUS-LC domain contains at least two distinct regions that could regulate the self-assembly of FUS: an N-terminal core-1 (residues 39-95) and a C-terminal core-3 (residues 155-190). (4) FUS-LC-C (residues 111-214) polymerizes faster than FUS-LC-N (residues 1-110). Conversely, thermal denaturation studies showed that the FUS-LC-N polymers are more stable, requiring higher temperatures for disassembly into soluble monomers, than the FUS-LC-C polymers. 

Here, we used coarse-grained simulations based on the Self-Organized Polymer for Intrinsically Disordered Proteins (SOP-IDP) model \cite{Baul19JPCB,Mugnai25ProtSci} to first demonstrate that the kinetics and stability of fibril-like structures are inversely correlated, which explains the experimental observation described above (item (1)). Second, multichain simulations show that the FUS-LC droplet forms by a nucleation and growth mechanism that is consistent with Ostwald ripening, in which the concentration of monomers and oligomers decreases at the expense of the growth of the dense droplet.  The coexisting concentrations between the dense and dilute phases, which we establish by showing the near-equality of the chemical potentials between the two phases, are in near-quantitative agreement with experiments. Third, the dense phase is initially driven by the formation of core-3.  SOP-IDP simulations show that key residues (164-176) adopt the $\beta$-strand conformation. The structure of core-3 fibril determined using AlphaFold shows prominent intra- and interchain contacts in this region. The combination of molecular simulations and structures derived from AlphaFold has resolved several puzzles in the FUS-LC assembly and sets the stage applications to other low-complexity IDP sequences.

\section*{Results}
\textbf{Ostwald's rule of stages:} 
To establish the relationship between the kinetics of formation of the three cores and their stabilites with respect to the random coil (RC), which is the ground state, we used the $N^*$ theory (summarized in the Supplementary Information) posits that within the native basin of attraction of the monomer high free-energy conformations are populated with low probability. In particular, the $N^*$ states resemble structural motifs found in amyloid fibrils \cite{straub2011toward,li2010factors,zhuravlev2014propensity}. 

The conformations belonging to the $N^*$ state are determined using long equilibrium simulations. We used the structural overlap function, $\chi$ (see Eq. 3 in the SI), describing the similarity between the conformation of the monomer and the structure in the fibril to characterize the excited states. For core-1 (core-2), we used fibril structure determined using   solid-state NMR \cite{core1Tycko} (PDB ID: 5W3N; see Fig. \ref{fig:AF3_core1andcore3}A) (cryo-EM \cite{core2Tycko} (PDB ID: 6XFM; see Fig. S15)). We used AlphaFold2 \cite{AF2} and AlphaFold3 \cite{AF3} to generate structures for the  prediction core-3 fibril (see Supplementary Results). We define the $N^*$ states as fibril-like conformations with structural overlap parameters $\chi_{fib.i} \geq \chi_{c.i}$, where $i = 1, 2, 3$ denotes the core. The procedure for selecting the values of $\chi_{c.i}$ is described in the SI Methods. The $N^{*}$-states are  populated in the conformational ensemble of monomers at extremely low levels (usually less than $\sim$1\%). Populations of core-1 and core-2  calculated using long equilibrium simulations showed \cite{kumar2021sequence,core1Tycko} that core-1 is only modestly more stable than core-2, which makes the absence of core-2 fibrils in FUS-LC \cite{core2Tycko} puzzling. Moreover, neither estimates of the time scales of formation of the fibril-like structures of the three cores nor the fate of core-3 has been investigated in experiments.

To establish a relation between the kinetics of formation of the three cores to their stabilities, we calculated the distribution of first passage times, $\displaystyle P(\tau_{FPT}) = \frac{1}{M} \sum_{i=1}^M \delta(\tau_{\mathrm{FPT}} - \tau_{1k})$, where $\tau_{FPT}$  is the first time at which the transition from the random coil (RC) ground state to the corresponding $N^{*}$ state occurs in the $k^{th}$ trajectory.   We associated $\tau_{\mathrm{FPT}}$ with the time  at which a structure with $\chi_{fib.i} \geq \chi_{c.i}$ is reached for the first time in various trajectories.  Using 200 independent Brownian dynamics simulations (see Methods), each lasting 280 ms, we observed 42 transitions to core-1, 87 transitions to core-2, 116 transitions to the U-bend conformation of core-3, and 124 transitions to the S-bend conformation of core-3. The smallest number of transitions to core-1 suggests that it is kinetically least accessible.  

Because  the first passage time (FPT) is not reached in all the trajectories, only truncated distributions of FPTs were sampled (Fig. \ref{fig:True_MFPT}B-D) in the simulations. Consequently, only the lower bounds for the true mean FPTs can be estimated. The error bars represent the standard error of the mean. We could conclude based on the calculated MFPTs that time for forming core-3 is the shortest while core-1 forms on the longest time. Because only a small fraction of the trajectories reach the $N^{*}$ states, there is uncertainty  in reaching this conclusion, which we resolve using a method described below. 

To assess the statistical significance of the differences in the mean first-passage times (MFPTs) to the various $N^*$ states, we conducted pairwise comparisons between the MFPT values using Welch’s unequal variances t-test \cite{welch1947}. The MFPT for core-1 ($129.17 \pm 11.73$ ms) is significantly greater than for core-2 ($97.47 \pm 8.70$ ms), the U-bend conformation of core-3 ($94.17 \pm 6.59$ ms), and the S-bend conformation of core-3 ($99.41 \pm 6.61$ ms), with $p$-values between 0.02 and 0.04. However, the differences between core-2, core-3 U-bend, and core-3 S-bend are not statistically significant ($p > 0.6$), nor is the difference between the two core-3 conformations. The lack of statistical significance is due to the finite duration of the simulations, which limits the sampling of rare transitions, resulting in underestimation of the MFPTs. To resolve this issue, we apply a correction described in Eq.~\ref{eq:true_MFPT} below.

 
 Although $P(\tau_{FPT})$, (Fig. \ref{fig:True_MFPT} (B-D)) is broad, the RC $\rightarrow  N^{*}$ transitions occur by two-state kinetics. The logarithm of the FPT distributions, $\ln[\tau_{FPT}]$, which could reveal kinetic traps as distinct modes \cite{woods2024analysis}, shows that the distributions are predominantly unimodal (Fig. \ref{fig:True_MFPT} (E-G)). Moreover, the fraction of trajectories which have not transited from RC~$\rightarrow~N^*$ transitions  as a function of time, computed as $P_{RC}(t) = 1-\int_0^t P(\tau_{FPT})\,d\tau_{FPT}$ also exhibit approximately exponential behavior (see Fig. S2F). Thus, we model the transitions using two-state kinetics, which allows us to approximate $P(\tau_{FPT})$, by an exponential distribution, $P(\tau_{FPT}) = \lambda_{i} e^{-\lambda_{i} \tau_{FPT}}$, where $\lambda_{i} = 1 / \widehat{\langle \tau_{c.i} \rangle}$ (Fig. \ref{fig:True_MFPT}H). We estimated the true MFPTs, $\widehat{\langle \tau_{c.i} \rangle}$ using Eq. \ref{eq:true_MFPT}. 
The precisely calculated MFPT values are $\widehat{\langle \tau_{c.1} \rangle} = $ 1554 $\pm $ 16 ms, $\widehat{\langle \tau_{c.2} \rangle} = $ 685 $\pm $ 10 ms, $\widehat{\langle \tau_{c.3U} \rangle} = $ 426 $\pm$ 7 ms, and $\widehat{\langle \tau_{c.3S} \rangle} = $ 404.9 $\pm$ 7.0 ms (see Fig. \ref{fig:True_MFPT}I).  These results unambiguously show that the order of formation of fibril-like structures is, core-3 followed by core-2 and finally core-1, $\widehat{\langle \tau_{c.1} \rangle} > \widehat{\langle \tau_{c.2} \rangle} > \widehat{\langle \tau_{c.3} \rangle}$. We should emphasize that MFPT estimates are only proxies for the times needed to form the fibrils, which occur on timescales of days in FUS-LC \cite{core1Tycko, core2Tycko, core3_McKnight, Berkeley21BJ}. 

We then assessed whether the $N^*$ states identified in the isolated monomer are structurally distinct from the rest of the conformational ensemble by comparing the residue-wise propensity for $\beta$-strand formation in fibril-like conformations—defined by $\chi_{fib.i} > \chi_c$ for each core $i = 1, 2, 3$—denoted as $p_{\beta}^{c.i}$, to that in the remainder of the ensemble, which predominantly consists of 
random-coil-like structures. Let  $p_{\beta}^{RC}$ be the residue-dependent $\beta$-strand content in fibril-like $N^*$ states (see Fig. S3). For each core $i$, the ratio $p_{\beta}^{c.i} / p_{\beta}^{RC}$ exceeds unity primarily within the corresponding core region of FUS-LC. This finding provides strong evidence that the $N^*$ states, although sampled by the isolated monomer, capture key structural features of fibril-like conformations. We also calculated the radius of gyration ($R_g$) and end-to-end distance ($R_{ee}$) for the $N^*$ states (see Fig. S4). The core-1 fibril-like conformations ($\chi_{fib.1} > \chi_c$) are more compact compared to both the random coil ensemble and the fibril-like conformations associated with core-2 and core-3 (see Fig. S4C). In contrast, core-2 and core-3 fibril-like conformations exhibit $R_g$ and $R_{ee}$ values that are more similar to those of the random coil states. 

Having shown that the excited states in the monomer ensemble are fibril-like structures,  we calculated the relative stabilities of the three $N^{*}$ states  (see SI Methods) sampled in the equilibrium Langevin simulations of FUS-LC.  
The free energy difference between the core-1 and core-2 is $\Delta F_{1,2} = -0.93 \pm 0.13$ kcal/mol, which is reasonably close to the value obtained from solubility measurements \cite{core2Tycko}, 0.35 $\pm$ 0.19 kcal/mol. The relative stability of core-2 with respect to the U-bend topology of core-3  is $\Delta F_{2,3U} = -7.37 \pm 0.15$ kcal/mol. Among the core-3 variants, the one with the S-bend topology is more stable than the U-bend conformation ($\Delta F_{3S,3U} = -5.59 \pm 0.12$ kcal/mol), while both are less stable than core-2 ($\Delta F_{2,3S} = -1.78 \pm 0.11$ kcal/mol). The stability hierarchy is  reflected in the energy differences $\Delta F_{1,3U} = -8.29 \pm 0.22$ kcal/mol and $\Delta F_{1,3S} = -2.70 \pm 0.10$ kcal/mol. Strikingly, the rates of formation of the three fibril-like structures are inversely correlated with with free energy differences between different cores  (see Fig. \ref{fig:True_MFPT}I), which is in accord with the Ostwald’s rule of stages, found previously in dipeptide supramolecular polymers~\cite{levin2014ostwald}.

\noindent \textbf{Absence of core-3 accelerates the formation of core-1 and core-2:} We repeated the calculations described above for FUS-LC-N (residues 1–163) to illustrate  the effect of core-3 on the kinetics of core-1 and core-2 formation. In 200  Brownian  trajectories, each lasting 60 ms, there are 16 transitions to core-1 and 100 transitions to core-2. The values of the MFPTs, estimated from the completed transitions are: $\langle \tau_{c.1|s} \rangle = 31.03 \pm 4.6 \, \text{ms}$, $\langle \tau_{c.2|s} \rangle = 27.41 \pm 1.7 \, \text{ms}$ (see Fig.  S2). The unimodal (Fig.  S2) distributions of the FPTs imply  $P(\tau_{\text{FPT}}) = \lambda_i e^{-\lambda_i \tau_{\text{FPT}}}$, allows us to calculate the true MFPTs, $\widehat{\langle \tau_{c.i} \rangle}$, using Eq. \ref{eq:true_MFPT}. We find that $\widehat{\langle \tau_{c.1} \rangle} = 694.97 \pm 0.01 \, $ ms, $\widehat{\langle \tau_{c.2} \rangle} = 85.14 \pm 0.002 \, $ ms (see Fig. \ref{fig:True_MFPT}J). Equilibrium simulations of FUS-LC-N yielded a free energy difference between core-1 and core-2 of $\Delta F_{1,2} = -2.24 \pm 0.06 $ kcal/mol. As is the case for FUS-LC, the formation rates of fibril-like structures and the free energy differences between the competing core structures in FUS-LC-N are inversely correlated (see Fig. \ref{fig:True_MFPT}J). Interestingly, core-1 and core-2 form 3-fold and 11-fold slower in FUS-LC than in FUS-LC-N, which lacks core-3 except for 9 residues (full core-3 consists of 36 residues). This finding is sufficient to conclude that core-3 must play a key role in the assembly of FUS-LC (however, see below).

\noindent \textbf{Droplet formation and Ostwald ripening:} 
We initiated five multi-chain simulations from five distinct initial conformations in a homogeneous solution containing 200 randomly distributed FUS-LC chains (refer to Methods; see also a top panel in Fig. \ref{fig:LLPS}A; see also Fig. S5). The box size corresponded to an overall protein concentration of 600 $\mu$M, which is larger than the typical range (100-250 $\mu$M \cite{yokosawa2022concentration, burke2015residue}) used in \textit{in vitro} experiments  but is sufficiently low to observe both the condensed and dilute phases in equilibrium (see Fig. \ref{fig:LLPS}A). 
\newline
The unbiased simulations show that oligomers of different sizes form spontaneously (see Supplementary Movie 1). We divided the oligomers into three groups: small oligomers (2-9 proteins), large oligomers (10-31 proteins), small droplets (32-70 proteins), and large droplets (> 70 proteins). At the start of the simulations ($t <0.1 \times 10^9\tau$, where $\tau$ is a simulation time step defined in Methods), the FUS-LC monomers quickly coalesce into small oligomers, which rapidly disintegrate into monomers and reassemble repeatedly. At intermediate times, some of the small oligomers act as nucleation sites, leading to the formation of large oligomers. The initial clustering is in accord with the expectations of the classical nucleation theory (CNT) for the formation of the subcritical nuclei \cite{Debenedetti96book}. Eventually, one of the oligomers serves as a nucleation site, leading to the spontaneous formation of a small droplet at $t >0.5 \times 10^9\tau$. The transition from a subcritical to a supercritical nucleus is a hallmark of CNT, signifying the instant when the nucleus exceeds the critical radius and initiates the stable growth of a large FUS droplet (see Figs. \ref{fig:LLPS} and S5). The onset time for stable droplet growth varies significantly across different simulation runs (see Fig. \ref{fig:LLPS}), suggesting that the droplet assembles by multiple pathways, which is also consistent with CNT predictions.

Most commonly, monomers disintegrate from the oligomers or smaller droplets, thus exposing  a higher fraction of sites to the solvent (modeled implicitly here). These are subsequently  reincorporated into larger droplets, which  is reminiscent of  Ostwald ripening \cite{sugimoto1978general,lifshitz1961kinetics}, a phenomenon in which less stable smaller droplets or oligomers disintegrate into monomers (see Figs. \ref{fig:LLPS}C and S9A; see also Supplementary Movie 2) due to their higher surface energy and chemical potential (see Fig. \ref{fig:contact_map}C). Such monomers subsequently diffuse and are re-incorporated into larger droplets or oligomers, which are  free energetically more favorable because of their lower surface-to-volume ratio and reduced interfacial energy. In our simulations, the higher fraction of solvent-exposed sites in smaller droplets (see Fig. S6) accelerates their dissolution, effectively supplying material for the growth of larger droplets, as anticipated by Ostwald ripening \cite{lifshitz1961kinetics,Wagner61Ostwald}. This process results in the coarsening of the system over time, with larger droplets becoming more prominent while smaller ones diminish. The observed dynamics highlight the interplay between thermodynamic stability and molecular exchange, which governs the temporal evolution of the phase-separated FUS-LC droplets. 

\noindent \textbf{Monomer conformations in the dilute and dense phase are similar but not identical:}  We compared the Small-Angle X-ray Scattering (SAXS) profiles ($I(q)$ with $q$ being the scattering wave vector) calculated for FUS-LC chains within the droplet to $I(q)$  in the dispersed phase (see Fig. \ref{fig:contact_map}A). The profiles were virtually indistinguishable from each other, which is consistent with several experimental studies on FUS-LC of varying sequence length. (1) In a recent work \cite{betasheetsOnSurface}, NMR signals for FUS NTD (residues 1-267) were indistinguishable between  the dilute and condensed phases. Similar conclusions were reached  using solid-state NMR studies of FUS-LC \cite{Tycko2024conformations}.  (2) Solution-state NMR studies of FUS-LC-N (residues 1-163) \cite{murthy2019molecular} showed no significant conformational changes upon phase separation on the timescale of the experiment. (3) Double electron-electron resonance (DEER) spectroscopy studies of FUS NTD, consisting of residues (1-267)  \cite{esteban2024ensemble} also showed that the conformational similarity of FUS in dilute and condensed phases. The measured distance distributions between selected pairs of residues within FUS NTD monomers inside and outside the condensed phase were similar \cite{esteban2024ensemble}. We calculated these distributions between the same set of residues in FUS-LC and found that they are similar both the phases (Fig. \ref{fig:dist_in_FUS267}). The averages and standard deviations of these distances show reasonable agreement with those reported for FUS NTD (see Table  S4; see also Fig. \ref{fig:dist_in_FUS267}). {
(4) The inferred distances between the  N-terminal residue and three selected residues in FUS-LC-N monomers both inside and outside the droplet using experiments smFRET \cite{joshi2023single} showed only a modest increase in the condensed phase.   
We computed the same distances for FUS-LC and similarly observed a minor increase for monomers inside droplets (see Fig. S11). This increase in distance between the N-terminal and other residues suggests a modest degree of monomer unwinding within the condensed phase, which is also reflected in the distributions of the radius of gyration ($R_g$) and  end-to-end distance ($R_{ee}$) for FUS-LC proteins inside and outside the droplet (see  Fig. S11). 

\noindent \textbf{Inherent Structures (IS):} The findings  described above show that while the conformations of monomeric FUS-LCs remain largely similar in homogeneous and phase-separated environments, subtle structural differences emerge in the latter. To further investigate this, we calculated the residue-dependent $\beta$-strand propensity of FUS-LC monomers in both the environments (see SI Methods)  (see Fig. \ref{fig:contact_map}D). The overall $\beta$-strand propensity remains low, suggesting that monomeric FUS-LCs inside droplets are also disordered. However, the monomers within droplets have nearly threefold higher propensity for $\beta$-strand formation compared to free monomers (see Fig. \ref{fig:contact_map}D). To ascertain that the strand content is enhanced in the droplet, we determined its inherent structure (IS) \cite{stillinger1984packing} using a method that was introduced to probe packing in dense liquids. The IS, which locates local minima in the energy landscape, was determined by quenching the conformation of the droplet to a low temperature. We find that the $\beta$-strand (see Fig.\ref{fig:contact_map}E) content is typically greater in the cores than elsewhere. In particular, the  greater $\beta$-strand content in core-3 shows that there is a higher tendency for structural ordering in this region even before core-1 forms, which is consistent with experimental observations \cite{core3_McKnight} and Ostwald's rule of stages (see Fig. \ref{fig:True_MFPT}). It is worth bearing in mind that fibrils form only on times scales on the order of days \cite{core1Tycko,core3_McKnight,Berkeley21BJ}, which implies that the droplets in our simulations and experiments \cite{murthy2019molecular} are metastable.

\noindent \textbf{Chemical potentials in the two phases are similar:}
For $t > 3.5 \times 10^9 \, \tau$, a stable droplet forms from which only the monomers disassociate on long time scales (see Fig. \ref{fig:LLPS}). However, polymer chains chains exhibit continuous dynamic exchange, moving in and out of the droplet (see Fig. \ref{fig:LLPS}B; see also Supplementary Movie 3), as required for phase equilibrium. To ensure that we characterize the equilibrium  state of the condensate, we discarded the first $3.5\times10^9\tau$ time steps in the trajectory. To assess that equilibrium is established, we calculated the chemical potentials in the condensed ($\mu_c$) and dispersed ($\mu_d$) phases using the approach described in SI. We assumed that the polymers that are not in the condensed phase constitute the dispersed phase. In other words, the variations in the oligomer sizes found in the dispersed phase are not treated as separate species. The relative difference, $\Delta \mu = (\langle\mu_c\rangle - \langle\mu_d\rangle)/\langle\mu_c\rangle$, is 0.003 ± 0.03. The large error bars that exceed the mean value are due the finite system size (see Fig. S7).  


The similarity between the chemical potential distributions of the two phases, $P_d(\mu)$ and $P_c(\mu)$ (see Fig. \ref{fig:contact_map}B), was assessed using the Jensen-Shannon  divergence (JSD) (see SI Methods). The calculated divergence, $D_{\text{JS}}(P_d(\mu) \parallel P_c(\mu))$ = 0.01. The small value of the JSD shows that the two distributions are similar. Because their mean values are comparable, we surmised that the observed differences  arise from the larger variance in $\mu$ for the condensed phase (see Fig. \ref{fig:contact_map}B). Given that the system size that can be reliably simulated is small (maximum of 200 chains),  the conclusion that the chemical potentials between the coexisting phases are similar is unreasonable.

\noindent \textbf{Concentrations of the coexisting phases:} The establishment of the equality of the chemical potentials in the two phases allows us to reliably calculate the coexisting protein concentrations (see SI Methods for details). The average FUS-LC concentration of the condensed phase is $\rho_c=(16.24\pm8\times10^{-4}$) mM (see, Figs. \ref{fig:LLPS}D and \ref{fig:contact_map}H; see also Fig. S12B), which is in excellent agreement with a few experiments. Raman imaging experiments were used to estimate the dense phase concentration to be 15.0 mM under standard conditions  (pH 7.6 and 150 mM NaCl) \cite{yokosawa2022concentration}. An independent study using quantitative phase imaging reported that the FUS-LC concentration in the condensed phase to be 337.3 mg/mL (15.7 mM, assuming a molecular weight of 21.5 kDa for FUS-LC) \cite{mccall2020quantitative}, which is comparable to the 20 mM concentration reported for FUS NTD (residues 1-267)  \cite{esteban2024ensemble}. The average FUS-LC concentration in the dispersed phase in the simulations is $\rho_d=(0.28\pm 2\times10^{-4})$ mM (see, Fig. \ref{fig:contact_map}H), which is in reasonable agreement with the experimentally reported concentration of $\approx$ 100 $\mu$M \cite{esteban2024ensemble, maltseva2023fibril}.  In our simulations we observed that on average there are $114.80\pm0.05$ FUS-LC chains in the condensed phase, while $85.2\pm0.05$ are in the dilute phase (see Figs. \ref{fig:LLPS}C and \ref{fig:contact_map}G), which agrees with experimental estimates \cite{esteban2024ensemble}. The elevated overall protein concentration used in the simulations likely results in the larger value $\rho_d$ of the low density phase. The density of the droplet calculated using simulations  is about $\approx$ 58 higher than in the dispersed phase. The corresponding experimental value is $\approx$ 157. The good agreement between the simulations and several experiments further validates our model. 

\noindent \textbf{Mechanism of droplet formation:} Given that the chains in the droplet are disordered, it is natural to wonder if specific interactions between the FUS-LC polymers initiate phase separation. To address this issue, we analyzed interchain interactions. The results, which were qualitatively consistent across all the simulation runs, are illustrated in Figs.  S8 and S9. At short times, interchain interactions are rare because most proteins exist as free monomers or small oligomers. However, the majority of interchain interactions that  form during the early times are mediated by the core-3 region, with frequent interactions between residues 164-176 of one protein and the corresponding residues of another. Subsequently, other non-preferential interactions, which are intermittent and disrupted over time, are observed. The contacts between residues 164-176 consistently form the most populated interchain interactions. Interactions mediated by core-1 are discernible only at later stages of droplet formation, with preferential contacts involving core-1 residues 61-77.  Interestingly, amino acid residues in core-2 (residues 112-150) do not engage in preferential interactions at either the beginning of the simulation (Figs.  S8 and S9) or not at later stages (Fig. \ref{fig:contact_map}I), except when mediated by core-1 or core-3. At equilibrium, interactions between core-2 regions of different chains are the least probable (see Fig. \ref{fig:contact_map}I), a finding that supports the observation that core-2  is not required to bind to hydrogels \cite{core3_McKnight}. 
\newline
These  findings were derived using simulations conducted at low friction,  which could alter the kinetic pathways and the temporal order of events. To mitigate these effects, we performed additional five independent Brownian dynamics simulations for a system of 64 FUS proteins in a solvent with viscosity set to $10^{-3}$ Pa·s (see Methods). In the first 150 $\mu$s of each simulation, transient oligomers containing 4-8 chains form and dissociate (see Fig. S9A), in accord  with the results from low-friction simulations (see Fig. \ref{fig:LLPS}). Analysis of the interchain interactions during Brownian dynamics further revealed that residues 164-176 primarily mediate  contacts between the chains in the oligomers (see Fig. S9). From the temporal progression of the contact maps (see Fig. S9C), we find that condensate formation is initiated from oligomers that act as nucleation sites, predominantly involving core-3. This suggests that core-3 plays a central role in mediating droplet formation (see below). This is further supported by the interaction patterns observed in dimers and small oligomers (fewer than 10 chains; see Fig. S9D): although the core-1 region also participates in interchain interactions, their relative contribution decreases in larger oligomers, where core-3-mediated interactions become dominant.

\noindent \textbf{Core 3 residues drive FUS-LC condensation:} It is surprising that the most frequent and the strongest interchain interactions are mediated by the C-terminal tail of FUC-LC, particularly residues \textsuperscript{164}SSGGGGGGGGGGN\textsuperscript{176} (see Fig. \ref{fig:contact_map}F,I). Strikingly, these residues have the highest propensity to form $\beta$-strands at the monomer level (see Figs. \ref{fig:contact_map}D and  S10). The relevance of core-3 in FUS-LC assembly was demonstrated  \cite{core3_McKnight}  by  systematically truncating residues from C terminus, and probing their ability to bind to  hydrogel. Truncation beyond residue 180 resulted in a complete loss of the ability of GFP-tagged test proteins to bind to mCherry hydrogels formed by the intact FUS-LC (see Fig. 2A in ref. \cite{core3_McKnight}), thus suggesting that  residues that are critical for FUS-LC condensate formation is within the 165-180 range. This finding is supported by  our simulations which show that the most probable interactions are mediated by residues 164-176. In addition,   our simulations show that residues  \textsuperscript{71}TPQGYGSTGGYGSS\textsuperscript{84} (Fig. \ref{fig:contact_map}F,I) also form several interchain interactions within the droplet. Kato and McKnight also systematically created FUS-LC variants by  truncating chunks of residues from the N terminus (see Fig. 2B in ref. \cite{core3_McKnight}). The results of these experiments show that the crucial residues for FUS-LC hydrogel binding activity are within residues 61-77, which also aligns with our simulation results. The early formation of structure in  residues  (164-176), containing mostly Glycines, the absence of discernible order in core-2, and emergence of order late in core-1 show that  core-3 that initiates the assemblly of FUS-LC.

\noindent \textbf{ Structural model for core-3:} 
In light of the importance of core-3 in FUS-LC assembly, we used to AlphaFold2 (AF2; \cite{AF2}) and AlphaFold3 (AF3; \cite{AF3}) to predict the unknown structure of core-3.   To assess the reliability of AF2 and AF3, we first determined the structures of core-1 and core-2 and compared them to the experimentally resolved counterparts (see Figs. \ref{fig:AF3_core1andcore3} and  S15 for AF3). AlphaFold3 generates structurally diverse models of the fibrils based on different random seeds \cite{AF3amyloids}. 
Although the predicted and experimentally derived structures deviate to some extent, the majority of residues that form $\beta$-sheet secondary structures in the experimental fibrils are also found in $\beta$-sheets in the AF3-predicted models (see Supplementary Results for details). Moreover, the predictions are sufficiently accurate to validate Ostwald's rule of stages. This conclusion is based on a comparison of mean first-passage times (MFPTs) computed using both AF3-predicted and experimental fibril structures (see Supplementary Results for details).

To assess the reliability of the AF3 predictions, we analyzed the quality metrics provided by AlphaFold—pLDDT, iPTM, and pTM \cite{AF3}. These metrics exhibited substantial variability depending on the chain configuration. By examining their distributions across different chain numbers (see SI for details), we determined the most reliable structural model (Fig.  S14). For FUS$_{155-190}$ fibrils, which consist of four subunits (Fig.  S15D), we observed higher pLDDT, iPTM, and pTM values compared to other assemblies (see Fig.  S14). The selected core-3 fibril model had the highest values across all three quality metrics: iPTM, pTM, and average pLDDT (see Table  S3). These values fall within the typical range observed for amyloid fibrils \cite{AF3amyloids}. A similar analysis for FUS$_{141-214}$, which forms core-3 fibrils experimentally \cite{core3_McKnight}, revealed that the optimal structural model also consisted of a four-subunit complex (Fig. \ref{fig:AF3_core1andcore3}C), with iPTM, pTM, and average pLDDT (see Table  S3). Given the experimental confirmation of core-3 fibril formation for 
this sequence, we adopted the structures derived for FUS$_{141-214}$ in further analysis.

AF3 predictions show that the two outer extended segments of core-3 correspond to residues 155–163 and 182–190, while the inner segment consists of residues 168–177 (Fig. \ref{fig:AF3_core1andcore3}C). Within this architecture, residues 156–163 and 172–190 form $\beta$-strands in a serpentine-like $\beta$-sheet arrangement, whereas residues 164–166 adopt a $\beta$-turn conformation that facilitates chain reversal. Residues 155 and 167–172 remain in a random coil conformation. Figures  S16A and  S16B display per-atom pLDDT scores for the FUS$_{155-190}$ and FUS$_{141-214}$ fibrils, respectively. The monomeric unit in the FUS$_{155-190}$ fibril adopts a U-bend topology with a "kinked" region spanning residues 176–184 (Fig.  S15D). In contrast, the core-3 region (residues 155–190) of the FUS$_{141-214}$ monomer adopts an S-bend topology (Fig. \ref{fig:AF3_core1andcore3}C). Interestingly, the "kinked" region (residues 176–184) is conserved across both models.  

Given that core-3 forms cross-$\beta$ polymers \cite{core3_McKnight}, we used the predicted structures to determine the  residue dependent secondary structural propensity \cite{pcasso}. Both the U-bend and S-bend topologies exhibited a $\beta$-sheet secondary structure throughout most residues (see Figs.  S16C,D). Additionally, residues 164-176, which showed strong inter-chain interactions in our simulations (Figs. \ref{fig:LLPS} and S9), also displayed prominent intra- and interchain contacts in the predicted core-3 fibrils (see Figs.  S16E,F). This result further highlights the importance of core-3 in mediating interactions critical for fibril assembly.

\section*{Discussion}
\noindent The in-depth study of phase separation in the low complexity domain of FUS and its variants has resolved a few long-standing problems related to its assembly.  For instance, based on the weak interactions between the polymer chains, one would expect that there are no preferential contacts between the chains within the droplet. 
Unexpectedly, our simulations reveal that  interchain interactions involving residues 164–176 (the center of the core-3) are structured  early and initiate the formation of the FUS-LC droplet, which implies that the presence of structure in core-3 propagates to core-1 over time.   
Importantly, the lack of interchain contacts associated with core-2  (Figs. \ref{fig:contact_map},  S8, and S9) supports the experimental finding  that its formation is not essential for hydrogel binding \cite{core3_McKnight}. Besides these conclusions, there are several other findings whose implications are explored below.

\noindent \textbf{Nucleation and Ostwald ripening:} Droplet formation and growth takes place in two steps. In the first step, monomers collide to form oligomers that frequently disintegrate until one with a critical size emerges (see Supplementary Movie 1), which subsequently grows to a larger droplet forms. In the second step, excess monomers (and possibly small oligomers) are ``devoured" (the word used in \cite{lifshitz1961kinetics}) by largest droplet leading to the coarsening of the condensate (see Supplementary Movie 2). The simulations vividly illustrate the two processes. 
During progressive coarsening  larger droplets become increasingly dominant (``winner take all" scenario \cite{li2008lattmodel}) as the smaller ones shrink.   The observed dynamics underscore the balance between thermodynamic stability and molecular exchange, which dictates the temporal evolution of the phase-separated FUS-LC droplets.

Our study provides  a structural perspective on the nucleation and growth of the droplet, bridging the initial nucleation events described by Classical Nucleation Theory (CNT) with the long-time coarsening governed by Ostwald ripening (see Supplementary Movie 1). Because of the structural heterogeneity of monomers and oligomers, nucleation mechanism may be most efficient only when a subset of structures of the heterogeneous conformations are involved. Indeed, the relevance of the heterogeneous structures sampled in the dispersed phase in controlling the droplet formation is supported by theoretical and experimental findings \cite{Bertrand23SciRep,Oranges24BJ}. Despite the complexity, remnants of CNT are evident in the droplet growth. Our simulations show that the dependence of droplet stability on size is in accord  with both theoretical frameworks: CNT establishes a critical nucleus size required for stable growth, while Ostwald ripening explains the preferential growth of larger droplets at the expense of smaller ones. The simulations capture this size-dependent behavior, elucidating its role in shaping the overall droplet population dynamics.



\noindent \textbf{Core-3 structure prediction:} To determine the kinetics of core-1 and core-2 formation using the spectrum of monomer states, we utilized  the  known experimental structures for reference  \cite{core1Tycko, core2Tycko}.  To determine the rate of formation  of core-3,  we relied on the structures predicted by AlphaFold. Although there is some variability among the structures, core-3 is predicted to have both a U-bend and S-bend topologies, with the same residues positioned at the kink. Using the structural overlap parameter, we calculated the mean first passage times and found them to be  similar for both types of structures. This shows the conclusions  pertaining to Ostwalds's rule of stages are robust even if alternative conformations predicted by AlphaFold are used (Fig.  S20). We do not claim that any of the predicted structures of core-3 are correct in reality. Indeed,  the performance of AlphaFold for IDRs and even for fold-switching proteins has been criticized \cite{ragonis2024can, AF3amyloids,Chakravarty24NatComm}. Despite these caveats, our results for FUS-LC are in excellent  agreement  with experiments \cite{core3_McKnight}, suggesting that predicted structural models for core-3 are likely to be  reasonable. 

\noindent \textbf{Ostwald's rule of stages:}  The detection of the excited $N^{*}$ fibril-like structures the spectrum of monomers \cite{kumar2021sequence} informs our understanding of fibrillar assemblies  in IDPs and proteins that have a high propensity to aggregate  \cite{patel2015liquid,zhuravlev2014propensity,neudecker2012structure}.   
The inverse correlation between the stability of the fibril cores and the kinetics of their formation, demonstrated from the knowledge of the $N^{*}$ states, is accord with the Ostwald rule of stages that was previously shown for dipeptide supramolecular polymers~\cite{Levin14NatComm} and A$\beta$ peptides~\cite{chakraborty2023Ostwald}.   Our simulations establish that determination of  the $N^{*}$ states, perhaps by relaxation dispersion nuclear magnetic resonance spectroscopy \cite{neudecker2012structure,Ceccon22PNAS} or by time resolved solid state NMR \cite{Jeon23NatComm}, would be important in anticipating the mechanism of condensate formation and subsequent transition to fibrils. It is worth noting that the validity of the Ostwald's rule of stages is shown using the structures of the monomer $N^{*}$ states. In actuality, FUS-LC fibrils form on time scales on the order of days \cite{core1Tycko,core3_McKnight,Berkeley21BJ}. Nevertheless, we expect that Ostwald's rule would hold as the droplets transition to a solid.

Although the $N^*$ conformations sampled by the isolated monomer are structural proxies for their conformation within core-1, core-2, or core-3 fibrils, our results show that these regions exhibit an increased propensity to form $\beta$-strand in the corresponding regions of FUS-LC (see Fig.  S3) even at the early stages of assembly. Thus, our findings validate the relevance of $N^*$ states in characterizing fibril formation even though they are sampled in isolated monomers but exhibit structural features that mirror those in the fibrils. This observation accords well with previous studies indicating that such $N^*$ states, though rarely populated under native conditions, are encoded within the monomer's free energy landscape and can serve as precursors to aggregation-prone conformations \cite{zhuravlev2014propensity}.

\noindent \textbf{Role of core-3:} The mutual exclusion in the formation of  core-1 and core-3  structures was established using ALS-associated mutations \cite{core3_McKnight}. Our simulations provide indirect evidence for this key finding.   It is based on a comparison of the MFPTs for fibril-like conformations in FUS-LC and FUS-LC-N (see Figs. \ref{fig:True_MFPT} and  S2). The absence of the core-3 facilitates the formation of core-1 and core-2, implying  that core-3 either energetically or through steric interactions hinders these transitions. The increased $\beta$-strand propensity in the core-1 region of FUS-LC-N compared to FUS-LC further shows that the presence of core-3 suppresses the cores with greater stability (see Fig.  S10). Importantly, the relative stability between core-1 and core-2, $\Delta F_{1,2}$, increases from $-0.93 \pm 0.13$ kcal/mol in FUS-LC to $-2.24 \pm 0.06$ kcal/mol in FUS-LC-N. The enhanced stability suggests that core-3 exerts a greater destabilizing effect on core-1 than on core-2, further supporting its role in modulating fibril formation pathways.  

\noindent \textbf{Implication for FUS-LC assembly:} Our results confirm that the thermodynamically more stable core-1 compared to core-3 ($\Delta F_{1,3S} = -2.70 \pm 0.10$ kcal/mol) forms significantly later than core-3 in FUS-LC (see Fig. \ref{fig:True_MFPT}I).    
Kinetic studies revealed that core-3 forms within 50 hours of incubation (see Fig. in Ref. \cite{core3_McKnight}). In contrast,  core-1 forms only after an incubation period of 5–7 days \cite{core1Tycko}. The combination of kinetic and thermodynamic findings suggests that core-3  must transition into core-1 over time.  Although  core-1 and core-3 do not  coexist in FUS-LC, our kinetic studies of FUS1-163 shows that the absence of the core-3 region facilitates the formation of core-1. The time to form fibril-like core-1 structure decreases by a factor of $\approx$4 in FUS-LC-N compared to FUS-LC (compare Figs. \ref{fig:True_MFPT} I and J).  This implies that, at the early stages of FUS-LC assembly, core-3 likely hinders core-1 formation. The mutual exclusivity of core-1 and core-3 was previously shown in experiments \cite{core3_McKnight} using ALS-associated mutations. 

Our simulations predict that in FUS-LC core-2 fibril-like structures form after  core-3 but before core-1.   Experiments demonstrated that core-2 is not required for hydrogel binding \cite{core3_McKnight}, which accords well with our simulations that  show that there are a limited number of interchain contacts between the core-2 regions in FUS-LC monomers within the condensates (see Fig. \ref{fig:contact_map}). The scarcity of favorable interactions likely impedes the formation of stable  core-2 fibrils.  The destabilization of core-2 could also arise from entropic repulsion between the ``fuzzy coats" surrounding the ordered regions of the non-polymorphic FUS-LS fibrils \cite{core1Tycko,core2Tycko}. 

Finally, there are similarities between  FUS-LS and the low complexity domain of the TAR DNA-binding protein 43 (TDP-
43). In the latter, core-1 (``aggregation" core) and core-2 (about 10 residues outside of core-1) transition from liquid to solid morphology \cite{Fonda21JACS}. The ordered cores do not typically coexist (as in FUS-LC) and form on different time scales \cite{Murray2025TDP}. It appears that Ostwald's rule of stages should be manifested in TDP-43 as well. The generality of the results discovered here (Ostwald's rule, nucleation and growth of condensates, maturation of droplets into fibrils an extremely long time scales) for other low complexity sequences remains to be established.    
\section*{Methods}
\noindent \textbf{Transition to  fibril-like monomer $N^*$ states:} Previously, we showed that the monomer ensemble has fingerprints of sparsely populated high free energy conformations with high degree of similarity to monomers in the fibrils \cite{kumar2021sequence}. Such conformations, belonging to the $N^*$ state, may be determined using long equilibrium simulations. The $N^*$ states were identified using the  structural overlap parameter (see Eq. (3) in the SI).    
\noindent To calculate the rates of transition to the $N^*$ states, we carried out Brownian dynamics simulations in the high friction regime, by setting the solvent viscosity of $10^{-3}$ Pa.s, which is roughly the water viscosity. The equation of motion of each bead $i$, derived by neglecting the inertial term in Eq. (2) in SI is, 
\begin{eqnarray}\label{eq:Brownian}
\dot{q}_{i} & = & -\frac{1}{\gamma}\frac{\partial U_{SOP-IDP}}{\partial q_{i}}+\Gamma_{i}.
\end{eqnarray}
\noindent We integrated Eq.\ref{eq:Brownian} using  the Euler algorithm in the custom LAMMPS code \cite{GShi2018interphase}. In the over-damped limit, the natural unit of time is $\tau_{HF}$ = $\frac{\gamma a^2}{k_{B}T}$, where $\gamma$ = $6\pi \eta a$, as described previously \cite{veitshans1997protein}. With $a$ = 2.88 \AA, the average bead radius in our model, we obtain $\tau_{HF}$ = 109.5 ps. To derive reliable estimates of the first passage times (FPTs), we generated $N_{sim} = $ 200 independent trajectories for each polymer at 298 K. The duration of each trajectories is 240 ms, which may not sufficient to  observe transitions to the $N^*$ states. To correct for this shortcoming  the procedure described below to obtain the true estimate of the transition times to the $N^*$ states.  

\noindent \textbf{True Mean FPT:} Because the duration of the simulation trajectories is finite the transition from the disordered ground  state to the $N^*$ state cannot be observed  in all the trajectories. As a result, the mean first passage time is underestimated.  To correct for the underestimate, we used the maximum likelihood estimate (MLE) to calculate the true mean FPT, by assuming that the first passage times FPTs are exponentially distributed. The true mean first passage time,  denoted as $\langle \tau \rangle$, is given by,
\begin{eqnarray}\label{eq:true_MFPT} \langle \tau \rangle & = & \frac{\sum_{k=1}^{N_{S}} \tau_{k} + N_{E} \cdot T}{N_{O}}.
\end{eqnarray}
\noindent The equation above  (Eq. \ref{eq:true_MFPT}), derived elsewhere \cite{shi2025static}  is similar to the approach in \cite{al2008estimation}, accounts for the contributions from events that occur within the duration ($T$) of the simulation trajectory as well those that do not on  $T$.  In Eq.~\ref{eq:true_MFPT}, $\tau_{k}$ is the FPT that is less than  $T$, $N_S$ is the number of trajectories,  $N_{E}$ denotes the number of simulations where the FPT exceeds $T$ (indicating cases where the first passage did not occur within $T$), and $N_{O}$ is the total number of observed FPTs that are less than $T$. The robustness of Eq.~\ref{eq:true_MFPT} was tested across different values of $N_S$ and $T$. It is worth pointing out that the validity of Eq. \ref{eq:true_MFPT} was established by explicit comparisons to precisely solvable models \cite{shi2025static}. 

\noindent \textbf{Multi-chain simulations:}  Each simulation run was initiated from different initial configuration of the system consisting of 200 polymers. The initial configurations were derived from five distinct conformations generated in the FUS-LC monomer simulations. The conformations correspond to a cluster in the ground state (see the Methods in SI). The dimensions of the monomers, $R_{g}$ and $R_{ee}$ values, were close to the ensemble average. Each of the five monomers was replicated 40 times and arranged in a simulation box using PACKMOL software \cite{packmol}, ensuring that there are no interactions between monomers in the initial configuration. The distance between the monomers in the initial configuration was 10-15 nm, which is larger than the average $R_{ee}$ and two-three times larger than the average $R_{g}$ of FUS-LC (Fig. S11). We use a temperature ($T = 209$ K) to facilitate droplet formation under equilibrium conditions. 

All the simulations were carried out at the salt concentration of 150 mM NaCl. To calculate the inherent structures \cite{stillinger1984packing}, we quenched the system to a low temperature, by isolating the droplet and then performed Langevin dynamics at $T = 3$ K until the potential energy converged.  Multi-chain simulations were performed on a GPU using OpenMM \cite{openmm7}.  The equations of motion were solved using the leap-frog integration scheme with a time step of $h = 0.05\tau$, where the natural time unit is, $\tau = a_0 \sqrt{m_0 / \varepsilon_0}$. Here, $m_0 = 1$ Da represents the mass scale, $a_0 = 1$~\AA\ is the characteristic length scale, and $\varepsilon_0 = 1$ kcal/mol sets the energy scale. To accelerate sampling, simulations were conducted in the underdamped regime by reducing 
the effective viscosity of water by a factor of 100. Each trajectory was propagated for a total of $5 \times 10^9$ integration steps.

\noindent \textbf{Clustering:} For each snapshot, a monomer was considered part of a cluster (i.e., an oligomer or droplet) if it was in contact with at least two other monomers within that cluster. Two monomers were defined to be in contact if there existed at least one pair of residues, $i$ and $j$, from the respective monomers such that the distance between their side chains was $\leq 0.8$ nm.

\noindent \textbf{Fibril structures:} To predict the three-dimensional structures of the fibrils, we employed AlphaFold2 (AF2; \cite{AF2}) through ColabFold, a Google-hosted AlphaFold collaborative Jupyter notebook that facilitates modeling of protein monomers as well as homo- and hetero-oligomeric complexes \cite{colabfold}. ColabFold was run with default parameters, with the number of subunits varied to assess the structural diversity. For AlphaFold3 (AF3; \cite{AF3}) predictions, we used the AlphaFold Server \cite{AF3} under default settings. To evaluate the reliability of each predicted structure, we analyzed multiple AlphaFold metrics: the Predicted local distance difference test (pLDDT), which reflects confidence in structure predictions; the  Template Modeling (pTM) score and the interface predicted template modeling (ipTM) score, which estimates the overall quality of the complex and the interfacial interactions within complexes, respectively \cite{AF2}. The pTM and ipTM scores approximate the Template Modeling (TM) score \cite{TMscore}, providing insights into the predicted structural accuracy. We present these metrics  for all the AlphaFold models in the SI Figs.  S14, S16,  S17.

\noindent \textbf{Acknowledgments:} We are grateful to Steve McKnight and Balaka Mondal for useful comments and critical reading of the manuscript. LD was partly supported by a Fulbright Fellowship. This work was supported by a grant from the National Science Foundation (CHE 2320256), and a grant from the Welch Foundation (F-0019) administered through the Collie-Welch Regents Chair. This work used Bridges-2 \cite{brown2021bridges} at Pittsburgh Supercomputing Center through allocations bio220054p and bio240056p from the Advanced Cyberinfrastructure Coordination Ecosystem: Services \& Support (ACCESS) program, which is supported by National Science Foundation grants \#2138259, \#2138286, \#2138307, \#2137603, and \#2138296. We acknowledge the Texas Advanced Computing Center (TACC) at The University of Texas at Austin for providing computational resources that have contributed to the research results reported within this paper. URL: \url{http://www.tacc.utexas.edu}.

\clearpage
\bibliographystyle{unsrt}
\bibliography{refs}
\clearpage
\begin{figure}
\centering
    \includegraphics[width=\textwidth,height=\textheight,keepaspectratio]{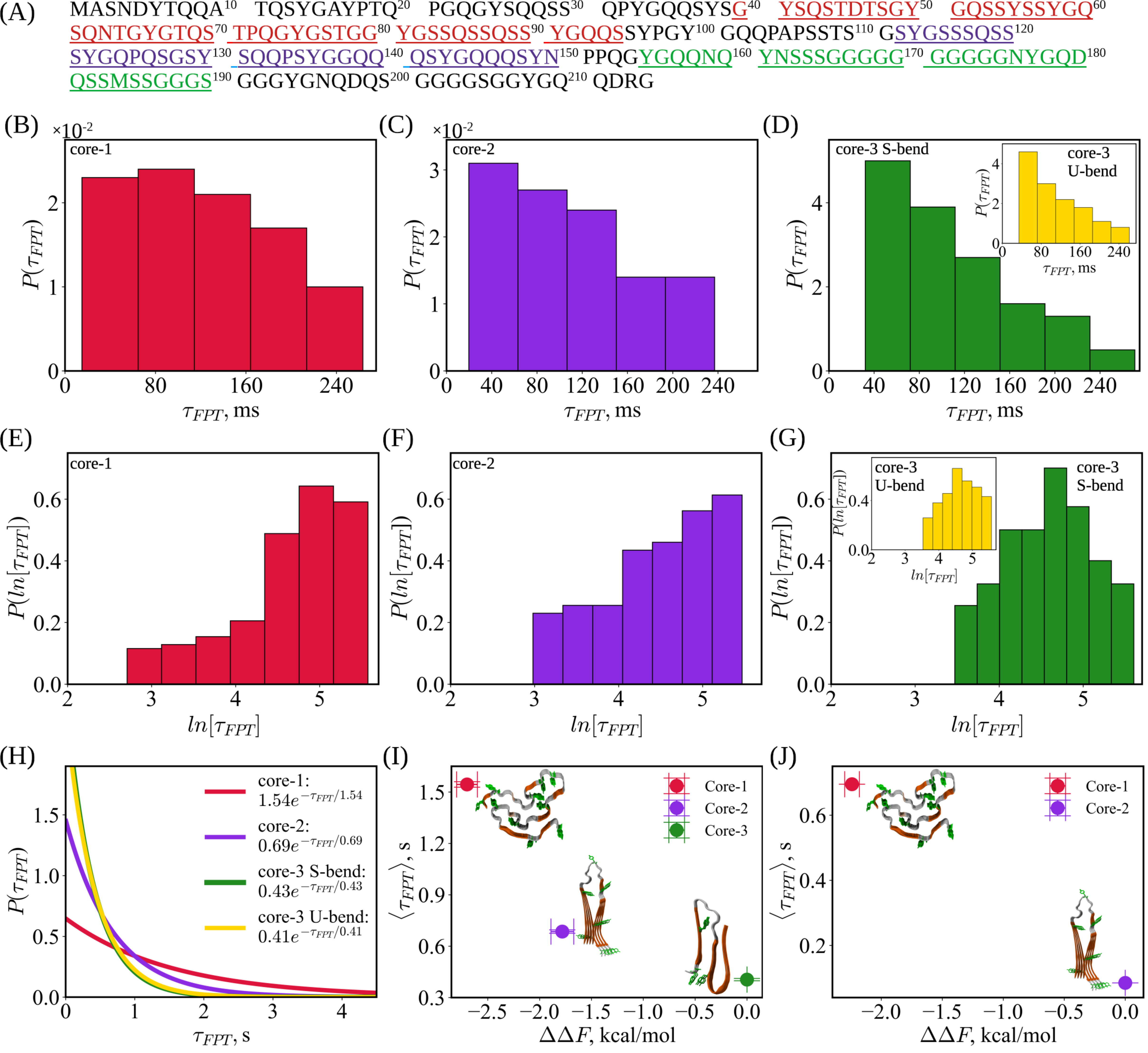}
    \caption{\textbf{Ostwald's rule of stages:}  (A) The FUS-LC sequence using a one-letter code for the amino acids. Core-1 residues 39-95 are  in red,  core-2 residues are highlighted in purple, and those forming core-3 are  in green. Panels B-D: Distributions of first-passage times (FPTs) for the transition from the disordered ground state to the fibril-like $N^{*}$ states in FUS-LC (residues 1–214). Shown are the $\tau_{FPT}$ distributions sampled from Brownian dynamics simulations (see Methods) for the transition from the free energy ground state of FUS-LC to the core-1 fibril-like structure (panel B), to the core-2 fibril-like structure (panel C), and to the core-3 S-bend and U-bend (inset) fibril-like structures (panel D) predicted by AF3. Panels E-G: Distributions of logarithm of first passage times (FPTs) $ln[\tau_{FPT}]$ for the transition from the disordered ground state to the fibril-like $N^{*}$ states. (H) Theoretical distributions of FPTs calculated using the correction formula (see Eq. \ref{eq:true_MFPT}). (I):  MFPTs as a function of relative stability with respect to core-3 (S-bend) for FUS-LC. The inset shows the experimental structures of core-1 (PDB ID: 5W3N \cite{core1Tycko}) and core-2 (PDB ID: 6XFM \cite{core2Tycko}), along with the predicted S-bend structure of core-3. (J) MFPTs as a function of relative stability with respect to core-2 for FUS-LC-N (residues 1–163). The distributions of $ln[\tau_{FPT}]$ for the transition from the disordered ground state to the fibril-like $N^{*}$ states for FUS-LC-N are show in Fig.  S2.}
\label{fig:True_MFPT}
\end{figure}
\clearpage

\begin{figure}[p]
\centering
    \includegraphics[width=\textwidth,height=\textheight,keepaspectratio]{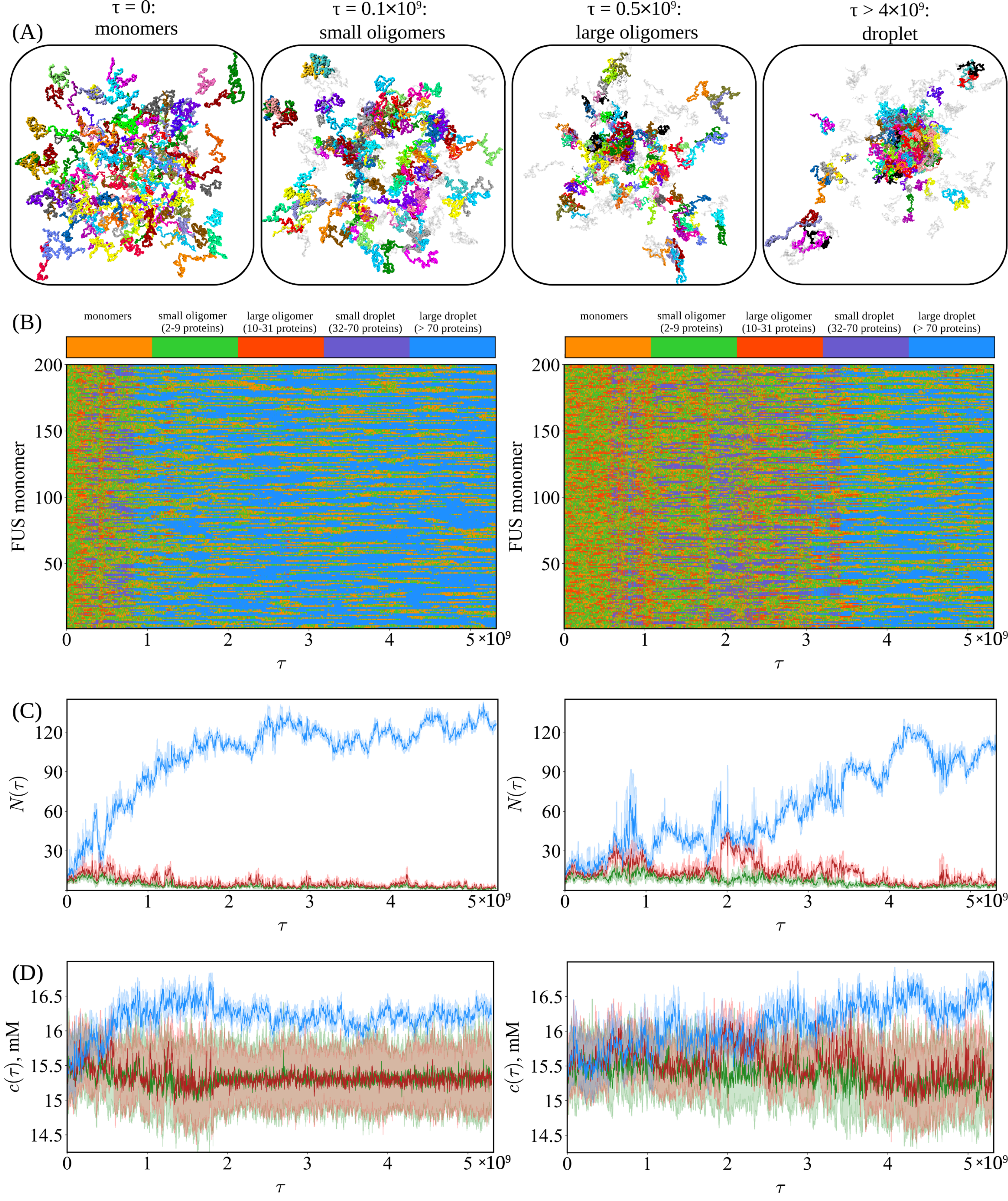}
    \label{fig:LLPS}
\end{figure}
\clearpage
\captionof{figure}{
\label{fig:LLPS}
\textbf{Phase separation in FUS-LC.}
(A) Representative snapshots illustrating the progression of phase separation in a system of 200 FUS-LC chains. The first panel shows the initial homogeneous configuration in the single dilute phase. Subsequent panels depict the emergence of small oligomeric nuclei, followed by their growth into a larger oligomer. This structure eventually matures into a droplet, representing the equilibrium state characterized by the coexistence of condensed and dilute phases. FUS-LC proteins in the monomeric state are rendered in transparent gray. (B) The temporal progression of individual FUS protein molecules is shown as a sequence of discrete states over the course of the simulation $\tau$ is a simulation time step, which is defined in Methods. Conformational states of the proteins are represented through a color-coded scheme. Two independent trajectories are displayed. (C)  Temporal evolution of the sizes of the three largest droplets (blue, red, and green) corresponding to the trajectories in panel B. Droplet size is defined as the number of FUS-LC chains contained within each droplet. (D) Protein mass concentration $c$ as a function of simulation time for the droplets shown in panel C. }

\begin{figure}[p]
\centering
    \includegraphics[width=\textwidth,height=\textheight,keepaspectratio]{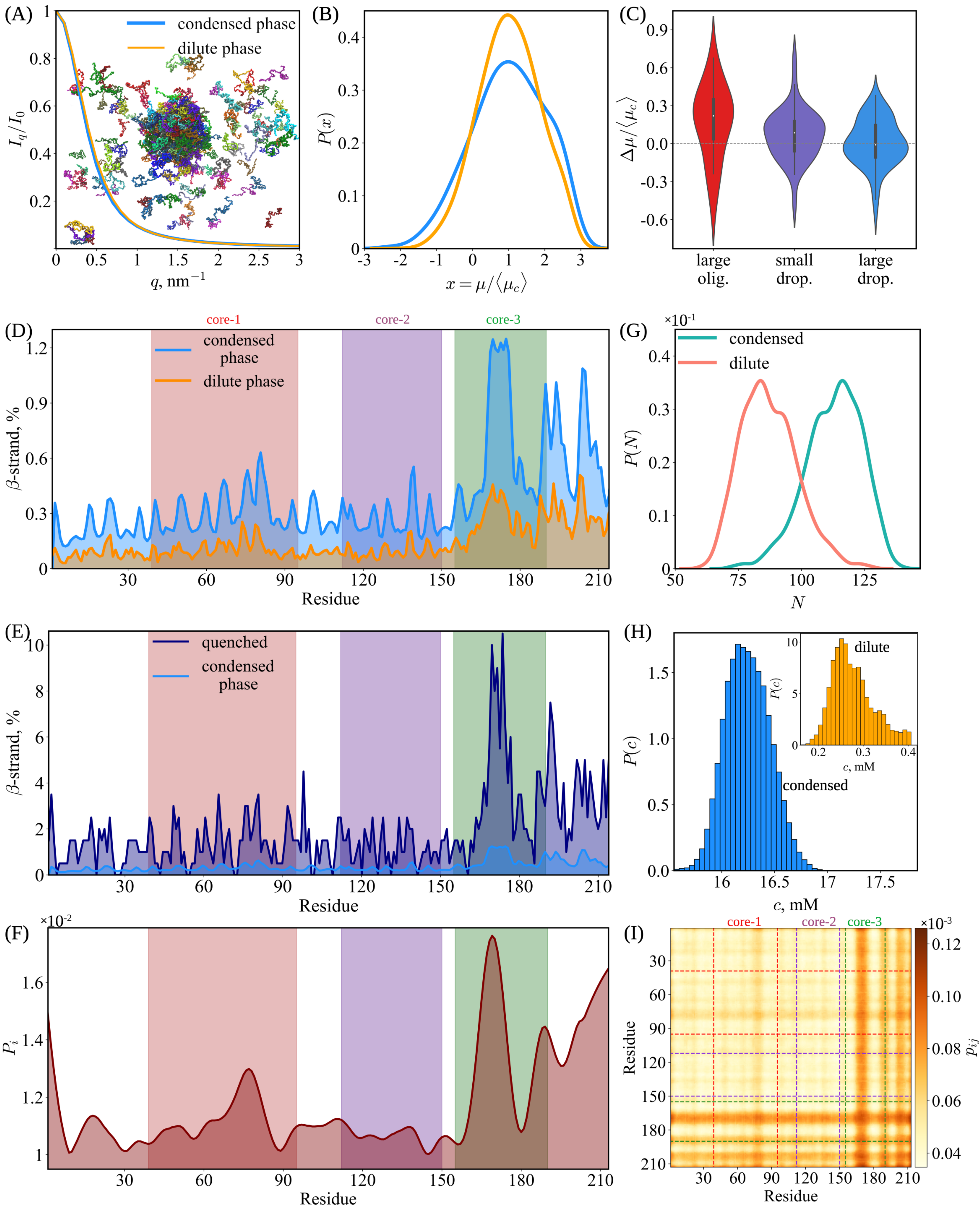}
    \label{fig:contact_map}
\end{figure}
\clearpage
\captionof{figure}{
\label{fig:contact_map}
\textbf{Equilibrium characteristics of FUS-LC condensates.}
(A) Small-angle X-ray scattering (SAXS) profiles for FUS-LC polymers in the condensed phase (within the phase separated droplets) and the dilute phase (outside droplets). The scattering intensity $I(q)$  as a function of the scattering vector $q$. The inset presents representative images from a simulation of 200 FUS-LC chains, illustrating the coexistence of dilute and condensed phases. (B) Distributions of the chemical potential in the dispersed and condensed phases, scaled by the average chemical potential in the condensed phase, $\langle \mu_c \rangle$. (C) Distributions of the chemical potential difference between condensed and dilute phases, defined as $\Delta \mu = \langle \mu_c \rangle - \langle \mu_d \rangle$, scaled by $\langle \mu_c \rangle$ for assemblies consisting of different numbers of chains, where $\langle \mu_c \rangle$ and $\langle \mu_d \rangle$ denote the average chemical potentials in the condensed and dilute phases, respectively. (D) Ensemble-averaged residue-wise percentages of $\beta$-strand content  for FUS-LC monomers in homogeneous and phase-separated environments. Vertical dashed lines are positioned at the residues corresponding  boundaries of the core regions: core-1 (residues 39-95), core-2 (residues 112-150), and core-3 (residues 155-190). (E) Amino acid dependent propensity for $\beta$-strand formation calculated from the inherent structure of droplet. The propensity values are averaged over all the chains in the condensate. (F) Cumulative probability for each residue of FUS-LC to form inter-chain contacts within the condensate. For a given residue $i$, $P_i$ is obtained by summing the average probabilities of all its intermolecular contacts with residues $j$ (see panel I). (G)  Droplet size distributions in the condensed and dilute phases. (H) Distributions of protein concentration in the condensed and dilute phases. (I) Intermolecular contact probability $p_{ij}$ between residues $i$ and $j$, calculated over the equilibrium portion of the trajectory ($t > 3 \times 10^9\tau$). }

\thispagestyle{empty}
\begin{figure}
\centering
    \includegraphics[width=\textwidth,height=\textheight,keepaspectratio]{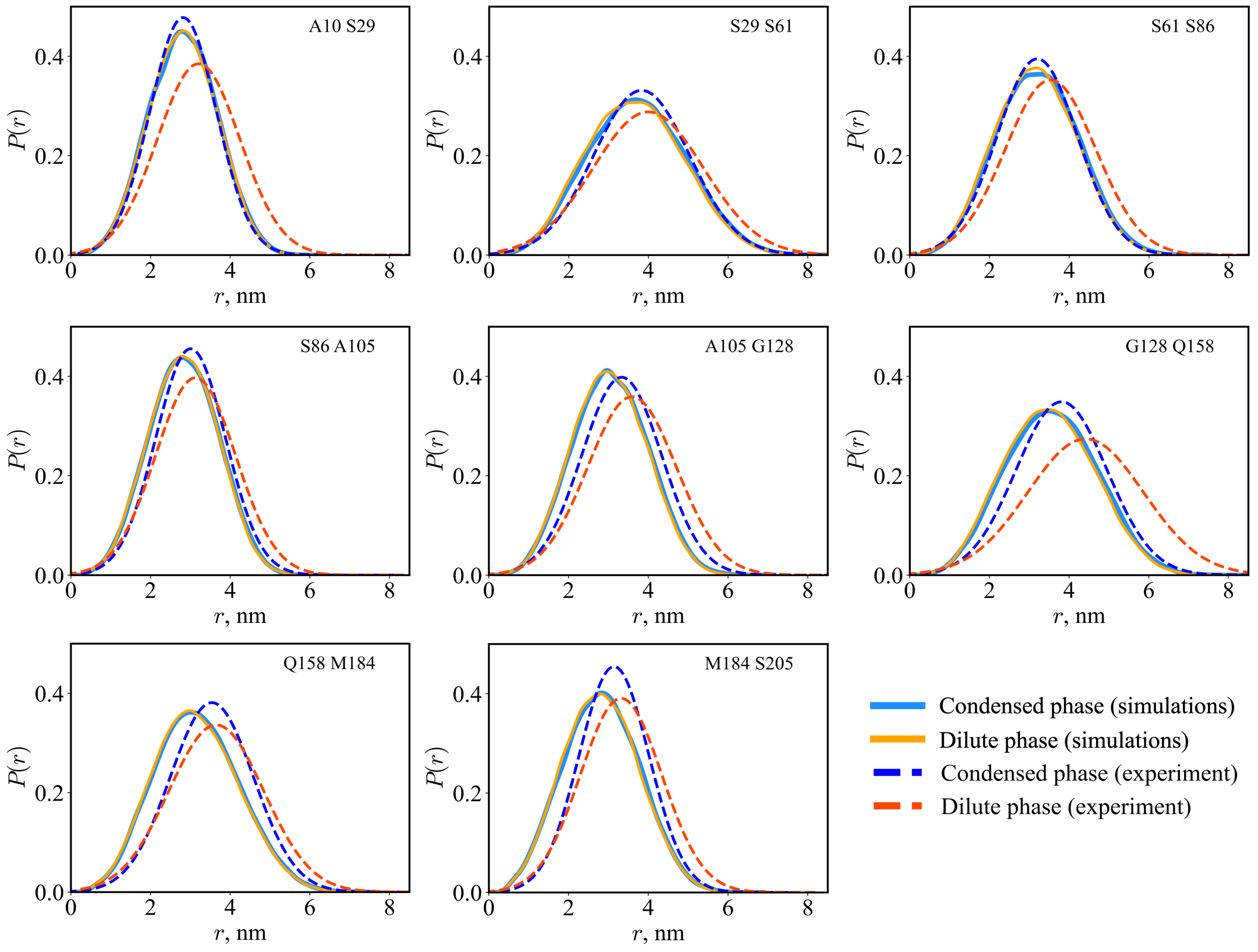}
    \caption{\textbf{Conformations of monomers in dilute and dense phase are similar.} Distance distributions between select residues (base on an experimental study \cite{esteban2024ensemble})  within the same FUS-LC chain  for chains in the dilute phase and those within the condensed droplet from simulations.  These distributions are compared to experimental distance distributions for FUS\textsubscript{1-267}, extracted from Fig. 2 in Ref. \cite{esteban2024ensemble}. The corresponding mean distances and standard deviations are provided in Table  S4.}
\label{fig:dist_in_FUS267}
\end{figure}
\clearpage


\begin{figure}
\centering
\includegraphics[width=\textwidth,height=\textheight,keepaspectratio]{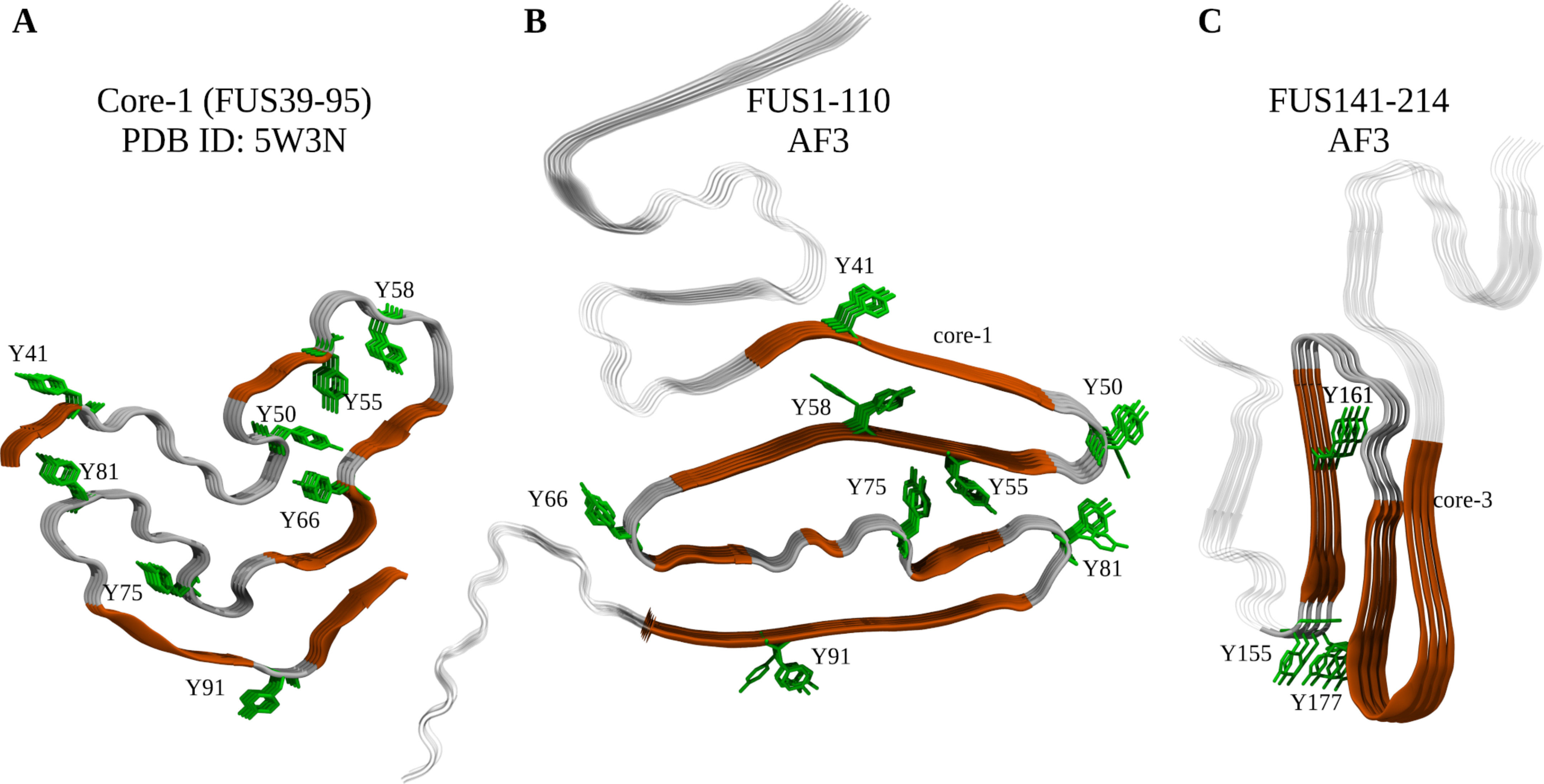}
\caption{\textbf{Structural models of core-1 and core-3 fibrils predicted by AlphaFold3.}  
(A) Structure of core-1 (residues 39--95) resolved by solid-state NMR \cite{core1Tycko}. Tyrosine (TYR) side chains are shown as sticks, with $\beta$-sheets colored brown and turn/coil regions in gray. (B, C) AlphaFold3 predictions for pentameric assembly of FUS\textsubscript{1-110} (B) and tetrameric assembly of FUS\textsubscript{141-214} (C). The color scheme is the same as in panel (A). Core-1 residues (39--95) in (B) and core-3 residues (155--190) in (C) are depicted as in panel (A), while the remaining fibril regions are rendered transparent for clarity.}
\label{fig:AF3_core1andcore3}
\end{figure}

\clearpage 

\end{document}